\documentclass[11pt]{article}

\usepackage{xspace} 
\usepackage[sort]{natbib}  % for bibliography; sort option makes the numbers in citations in a sensible order
\usepackage{hyperref}

\bibpunct{[}{]}{,}{n}{,}{,}  % controls how citations appear in text
\title {Software Abstractions and Methodologies for HPC Simulation
  Codes on Future Architectures}
\author{A. Dubey, S. Brandt,  R. Brower, M. Giles,  P. Hovland, \\
D.Q. Lamb, F. L\"{o}ffler,  B. Norris, B, O'Shea, C. Rebbi, M. Snir, R. Thakur}
  
\begin{document}

% superscripts and subscripts in text (for C02) see http://anthony.liekens.net/index.php/LaTeX/SubscriptAndSuperscriptInTextMode

%-----------------------------------------------------------------------------

\maketitle

Large, complex, multi-scale, multi-physics simulation codes, running on
high performance computing (HPC) platforms, have become essential to
advancing science and engineering. Progress in computational science,
together with the adoption of high-level frameworks and modular
approaches, have enabled large code development efforts such as FLASH 
\citep{Flash2010, Dubey2009, Fryxell2000}, Cactus \citep{Allen2000,Cactuscode:web},
Enzo \citep{O'Shea2005, Norman2008} and the Lattice QCD code suite 
\citep{QCDChroma,QCDquda,QCDsw}. These codes simulate multi-scale,
multi-physics phenomena with unprecedented fidelity on petascale
platforms, and are used by large communities. Increasing the
capabilities of these codes and maintaining their ability to run on
future platforms are as crucial to these communities as continued
improvements in instruments and facilities are to experimental
scientists. However, the ability of code developers to do these things
faces a serious challenge with the paradigm shift underway in platform
architecture from coarse grained parallelism with few highly capable
processors, to fine grained parallelism with many less capable and
more  heterogeneous processors. The complexity of the future platforms 
makes it essential to approach this challenge cooperatively as a
community. We need to develop common abstractions, frameworks,
programming models and software development methodologies that can be
applied across a broad range of complex simulation codes, and
common software infrastructure to support them. We believe that such
an infrastructure is critical to the deployment of existing and new
large, multi-scale, multi-physics codes on future HPC
platforms. Furthermore, such an infrastructure should be assembled by
a collaborative effort of the teams that develop and maintain such
codes -- that is, by the people that own the problem, not by people that
own solutions looking for a problem.   

% Effective utilization of high performance computing (HPC)
% resources has always been a balancing act between  portability and
% performance. Expected heterogeneity and ongoing deep architecture
% changes in the HPC platforms from one generation to the next move this
% balancing act beyond the resources of most code teams. Several
% individual efforts are underway, but they are fragmented and are
% therefore unlikely to lead to long-term sustainable
% solutions. Additionally, there will inevitably be duplications of 
% ideas and efforts if this piecemeal approach continues. For example,
% one common theme being explored by several groups is the use of domain
% specific languages (DSL) as a way of hiding some of the architecture
% complexity from the solvers. Each domain has its own preferred features
% in their DSL, which could potentially lead to an explosion in the
% number of DSLs and DSL construction tools that do not differ
% significantly from one another. One can hope that DSLs can be built
% using a common infrastructure and that there can be significant reuse
% of common constructs. 

The needs of the expert programmers who are developing complex,
high-performance simulation codes are quite different than the needs
of the broader community.  Furthermore, the number of such software
developers is small; their needs are therefore often ignored.  There
is need to foster cooperation among the small number of
teams that develop such codes, leading to a more viable ecosystem; and
cooperation of these teams with computer science researchers and
vendors, leading to a better understanding of the needs of this small
community.  Also, there is usually a gap between computer science
research in  areas such as code abstractions and transformations and
high-level scientific application software. The gap exists for many
reasons, including lack of communication between the communities
involved, limits on  extensibility and adaptability of scientific
software, and differing real and perceived requirements by different
application domains. It would be beneficial to the applications
community with HPC to overcome these and other barriers to finding and
implementing broad, reusable solutions and common software
infrastructure to support the deployment of existing and new
multi-scale, multi-physics codes on future HPC platforms. This would
also be a huge leap forward for the concerned science domains.  

A similar transformation took place in the 1990s when the groups
developing large scientific codes recognized the need for adopting 
software engineering practices to sustain code reliability in face of rapid
capability growth. Each group went its own way and customized some of
the prevalent ideas for its own use, many times developing their own
tools. However, a close examination of the adopted practices reveals a
surprising number of commonalities; object-oriented design, version
control, coding standards, unit testing, regression testing,
verification frameworks, release policies, contribution policies
etc. A shared infrastructure would not only have reduced development
costs, but would have facilitated code sharing.

To understand the kind of change software must undergo to adapt to new
computing platforms, we first consider the current characteristics of
the codes in our target community. They are typically implemented in
one or more of C, C++, or Fortran. Parallel computations are
implemented primarily using MPI, OpenMP, or a hybrid
approach. Underlying the particular simulation implementations are a
variety of mesh types, with or without adaptive refinement support,
and a number of implicit and explicit solvers with one or more
implementations of each method. Application-specific (I/O-intensive)
checkpointing schemes are typically used for fault tolerance and to
provide restart capabilities. Some simulation codes rely on externally
developed numerical libraries, while others are mostly self-contained,
with few external dependencies.  

The report of the 2011 Workshop on Exascale Programming
Challenges~\cite{ExascaleChallengesReport2011} discusses in depth many
of the challenges that scientific software is facing in the near
future, ranging from programming models to runtime systems. Many of
these challenges are not limited to extreme scales and are present even now
in HPC codes as they are faced with ever increasing levels of
parallelism and heterogeneity at any scale. We briefly overview
these software challenges

\begin {itemize}
\item {\bf Numerical methods and frameworks.} 
Certain disruptive architecture changes may require rethinking of
numerical approaches. Identifying the methods critical to our target
communities will be critical in assessing the applied mathematics
research and development required to effectively use future
platforms. A number of numerical \emph{frameworks} provide a means of
integrating a number of solution methods and underlying data
structures. Adapting the framework to new platforms is one way in
which multiple applications can successfully migrate to new paradigms
without significant reimplementation of application code. 

\item{\bf Programming Models.}
The programming model (e.g., MPI) used in a particular application and
its supporting infrastructure is a fundamental, pervasive aspect of
the implementation. Switching between programming models is
labor-intensive and may require significant redesign of key
algorithms, as well as massive code rewrites. At the same time, it is
difficult to estimate a priori the benefits of moving to a different
programming model, for example, when switching from a pure distributed
memory MPI-based implementation to a model that supports a global
shared memory view.  

\item{\bf Performance Portability.}
Effective utilization of HPC resources is historically a balancing act
between portability and performance. On one hand, extensive
performance optimizations of certain key computations are crucial for
achieving good performance on a certain platform. On the other hand,
making such changes permanent negatively impacts code readability and
performance on other platforms. it is important to 
identify and eliminate barriers to enabling greater flexibility in choosing
among different optimized implementations. Another critical aspect of
future performance portability is the ability to support different
levels and types of parallelism.  

\item{\bf Resilience.}
Current codes' reliance on checkpointing for error correction will be
prohibitively expensive on large-scale systems; hence, new ways of
ensuring resilience are required, which may involve changes in
programming models, runtimes, and application design and
implementation. 

\item{\bf Productivity and Maintainability.} 
Alternative technical approaches to managing increased levels of
parallelism, heterogeneity and other architectural features have
different impact on programmer effort and software
maintainability. Furthermore, complex codes do not allow easy testing
of new concepts and thus slow down advances in both the scientific and
numerical approaches. We must understand the current and desired mode
of development in our target communities to guide the approach to
design decisions.

\end{itemize}

It is a given that the changes will be disruptive, but the degree of
disruption can be reduced by working together as a community. We are
working on conceptualization of a Software Institute to address these
concerns, and are conducting workshops under the project that bring
together domain experts in several scientific fields and manufacturing
sectors, software developers  who are involved in implementing and
optimizing many of the large codes for these fields and sectors,
researchers in applied computer science, and hardware and software
vendors in a series of workshops and focus groups to gather a variety
of perspectives and broad expertise. The goal of the workshops has
been, and will continue to be defining and developing possible
approaches toward common abstractions and frameworks. 

This exercise can potentially produce a three-fold benefit to
the S2I2 program and to the community: (1) several large scientific
and engineering codes ready for and adaptable to generations of
heterogeneous and many-core platforms; (2) a common software
infrastructure that is applicable across a broad range of
science and engineering application domains; and (3) a model for
interaction between computer science research and application
development that takes interesting and promising research ideas from
simplified problems to real world applications.
%\input {proposal/journal_abbreviations}
%\bibliographystyle{plainnat}
%\bibliography{proposal/proposal,proposal/einsteintoolkit}

\end{document}